\documentclass[%
reprint,
superscriptaddress,
 amsmath,amssymb,
 aps,
floatfix,
]{revtex4-1}

\usepackage{graphicx}
\usepackage{dcolumn}
\usepackage{bm}
\usepackage{hyperref}
\usepackage{subscript}
\usepackage{epstopdf}
\usepackage{xcolor}
\hypersetup{
    colorlinks,
    linkcolor={red!50!black},
    citecolor={blue!70!black},
    urlcolor={blue!70!black}
}

\begin{document}


\title{Magnetic domain texture and the Dzyaloshinskii-Moriya interaction in 
Pt/Co/IrMn and Pt/Co/FeMn thin films with perpendicular exchange bias}

\author{Risalat A. Khan}
\email[Correspondence: ]{R.A.Khan@leeds.ac.uk}
\email[\\*Official contribution of the National Institute of Standards and Technology; not subject to copyright in the United States. ]{}
\affiliation{School of Physics and Astronomy, University of Leeds, Leeds LS2 9JT, UK}

\author{Hans T. Nembach}
\affiliation{Quantum Electromagnetics Division, National Institute of Standards and Technology, Boulder, Colorado 80305, USA}

\author{Mannan Ali}
\affiliation{School of Physics and Astronomy, University of Leeds, Leeds LS2 9JT, UK}

\author{Justin M. Shaw}
\affiliation{Quantum Electromagnetics Division, National Institute of Standards and Technology, Boulder, Colorado 80305, USA}

\author{Christopher H. Marrows}
\affiliation{School of Physics and Astronomy, University of Leeds, Leeds LS2 9JT, UK}

\author{Thomas A. Moore}
\affiliation{School of Physics and Astronomy, University of Leeds, Leeds LS2 9JT, UK}

\date{\today}

\begin{abstract}

Antiferromagnetic materials present us with rich and exciting physics,
which we can exploit to open new avenues in spintronic device applications.
We explore perpendicularly magnetized exchange biased systems of Pt/Co/IrMn
and Pt/Co/FeMn, where the crossover from paramagnetic to antiferromagnetic
behavior in the IrMn and FeMn layers is accessed by varying the thickness.
We demonstrate, through magneto-optical imaging, that the magnetic
domain morphology of the ferromagnetic Co layer is influenced by the
N{\'e}el order of the antiferromagnet (AFM) layers. We relate these variations
to the anisotropy energy of the AFM layer and the ferromagnet-antiferromagnet (FM-AFM)
inter-layer exchange coupling. We also quantify the interfacial Dzyaloshinskii-Moriya
interaction (DMI) in these systems by Brillouin light scattering spectroscopy.
The DMI remains unchanged, within experimental uncertainty,
for different phases of the AFM layers, which allows us to conclude
that the DMI is largely insensitive to both AFM spin order and exchange
bias. Understanding such fundamental mechanisms is crucial for the
development of future devices employing chiral spin textures, such
as N{\'e}el domain walls and skyrmions, in FM-AFM heterostructures.


\end{abstract}

\maketitle


\section{Introduction}

The field of spintronics \cite{vzutic2004spintronics} aims to realize
low-power and high-performance next-generation memory \cite{parkin2008magnetic}
and logic devices \cite{allwood2005magnetic,fukami2009low} through
the manipulation of the electron spin. Influencing ferromagnet (FM)
spins using an antiferromagnet (AFM) is an emerging branch of spintronics
\cite{duine2011spintronics,fukami2016magnetization,oh2016field,lau2016spin,tshitoyan2015electrical}.
The magnetization in a FM layer can be controlled by an adjacent AFM
layer through the interfacial coupling between the two layers \cite{meiklejohn1956new}.
AFMs have several other advantages as well. For instance, the net
magnetization is zero due to the compensation of magnetic moments
at the atomic level. The elimination of stray fields could prove to be vital
in integrated devices with low dimensions because such parasitic fields
(e.g., from a FM) present complications, such as crosstalk between
neighboring devices, susceptibility to external magnetic fields, etc.
Furthermore, AFMs possess excellent magneto-transport properties which
would allow the generation of large spin currents through which magnetization
in an adjacent FM layer could be efficiently switched \cite{tshitoyan2015electrical,fukami2016magnetization,zhang2016giant}.
AFMs also offer dynamics in the terahertz range suitable for ultrafast
information processing \cite{satoh2010spin}.

The exchange interaction is at the heart of magnetic behavior in materials.
It comprises a symmetric and an antisymmetric term. The symmetric
term, the Heisenberg interaction, prefers collinear orientation of
adjacent spins. The antisymmetric term, the Dzyaloshinskii-Moriya
interaction (DMI) \cite{dzyaloshinsky1958thermodynamic,moriya1960anisotropic},
prefers canted orientation of neighboring spins. In order to exist,
the DMI needs spin-orbit interaction in an asymmetric
crystal field, such as in heterostructures lacking spatial inversion
symmetry. The DMI gives rise to chiral spin textures \cite{bode2007chiral,meckler2009real},
which results in many different interesting phenomena \cite{ryu2013chiral,schulz2012emergent}.
In ultrathin film multilayers, the DMI is of the interfacial
form and has been reported to be present at the heavy-metal/ferromagnet (HM/FM)
interface \cite{emori2013current,ryu2013chiral}, at the FM/oxide
interface \cite{boulle2016room,belabbes2016oxygen}, and more recently,
at the FM/AFM interface \cite{ma2017dzyaloshinskii}. The DMI stabilizes
spin structures such as chiral N{\'e}el domain walls (DWs) \cite{thiaville2012dynamics}
and skyrmions \cite{fert2013skyrmions}, both of which can be driven
as information carriers \cite{parkin2008magnetic,fert2017magnetic}
by electric currents via the spin Hall torque generated in an adjacent
HM \cite{liu2012spin} and/or AFM \cite{tshitoyan2015electrical}
layer.

In this work, we investigate ultrathin film systems of Pt/Co/IrMn
and Pt/Co/FeMn, which exhibit perpendicular magnetic anisotropy (PMA)
and perpendicular exchange bias (PEB) \cite{maat2001perpendicular,garcia2002exchange,marrows2003three,sort2005tailoring}.
These multilayers are potentially of interest because of the coincidence
of the DMI with a vertical exchange field that could substitute the
need for an externally applied field to stabilize skyrmion bubbles
\cite{moreau2016additive}. We explore the interaction mechanisms
at the interfaces, in particular the changes in the magnetic domain
texture and the DMI, when going through the paramagnet to AFM phase
transition of the AFM layers by systematically varying the thickness
of the layers. When the AFM layer is in the paramagnetic
phase, the domains of the FM layer are large and contain networks
of unreversed narrow domains. As antiferromagnetic order sets in,
bubble domains with smooth DWs are nucleated. The DWs eventually become
rough at the onset of the exchange bias field. The nucleation density
also increases significantly. We relate this variation in the domain
morphology to the interplay between the anisotropy energy of the AFM
layer and the exchange energy at the interface between the FM and
the AFM layers. We identify the N{\'e}el and the blocking temperature
of IrMn to confirm paramagnetic behavior at low layer thicknesses.
We do this by exploiting the previously shown fact \cite{ali2003antiferromagnetic}
that these temperatures can be tuned by varying the AFM layer thickness.
Finally, we evaluate the interfacial DMI in these systems by Brillouin
light scattering (BLS) \cite{nembach2015linear,moon2013spin,di2015direct}.
We measure the DMI at four different phases of the AFM layer: paramagnet
phase, AFM phase without exchange bias (EB), AFM phase at the onset
of EB, and AFM phase with a large EB. The DMI is similar
for all four phases, from which we conclude that there is little influence
of AFM spin order or EB on the DMI in these systems. Investigating
such interactions provide insight towards the development of future
DW and skyrmion devices incorporating an FM-AFM bilayer.

\section{Multilayer systems}

The material systems that we studied consist of Pt(2 nm)/Co(1 nm)/Ir\textsubscript{20}Mn\textsubscript{80}(\emph{t}\textsubscript{IrMn})
and Pt(2 nm)/Co(0.6 nm)/Fe\textsubscript{50}Mn\textsubscript{50}(\emph{t}\textsubscript{FeMn})
trilayers deposited on a 5 nm Ta seed layer on a thermally oxidized
Si substrate. The Ta seed layer provides a (111) texture for the Pt
and Co layers, and consequently, for the IrMn and FeMn layers. Such
a crystal orientation is required for IrMn \cite{aley2008texture}
and FeMn \cite{jungblut1994orientational} for an effective exchange
coupling leading to a large EB. The layers were grown by dc magnetron
sputtering at a base pressure of $3\times10^{-6}$ Pa ($2\times10^{-8}$ Torr) and at an
Ar working pressure of 0.33 Pa (2.5 mTorr). A 3 nm capping layer
of Pt or Ta was also deposited on top of the stacks in order to prevent
oxidation. A change in capping layer has no effect on the magnetic
properties we measure here. 

\begin{figure}
\begin{centering}
\includegraphics[scale=0.32]{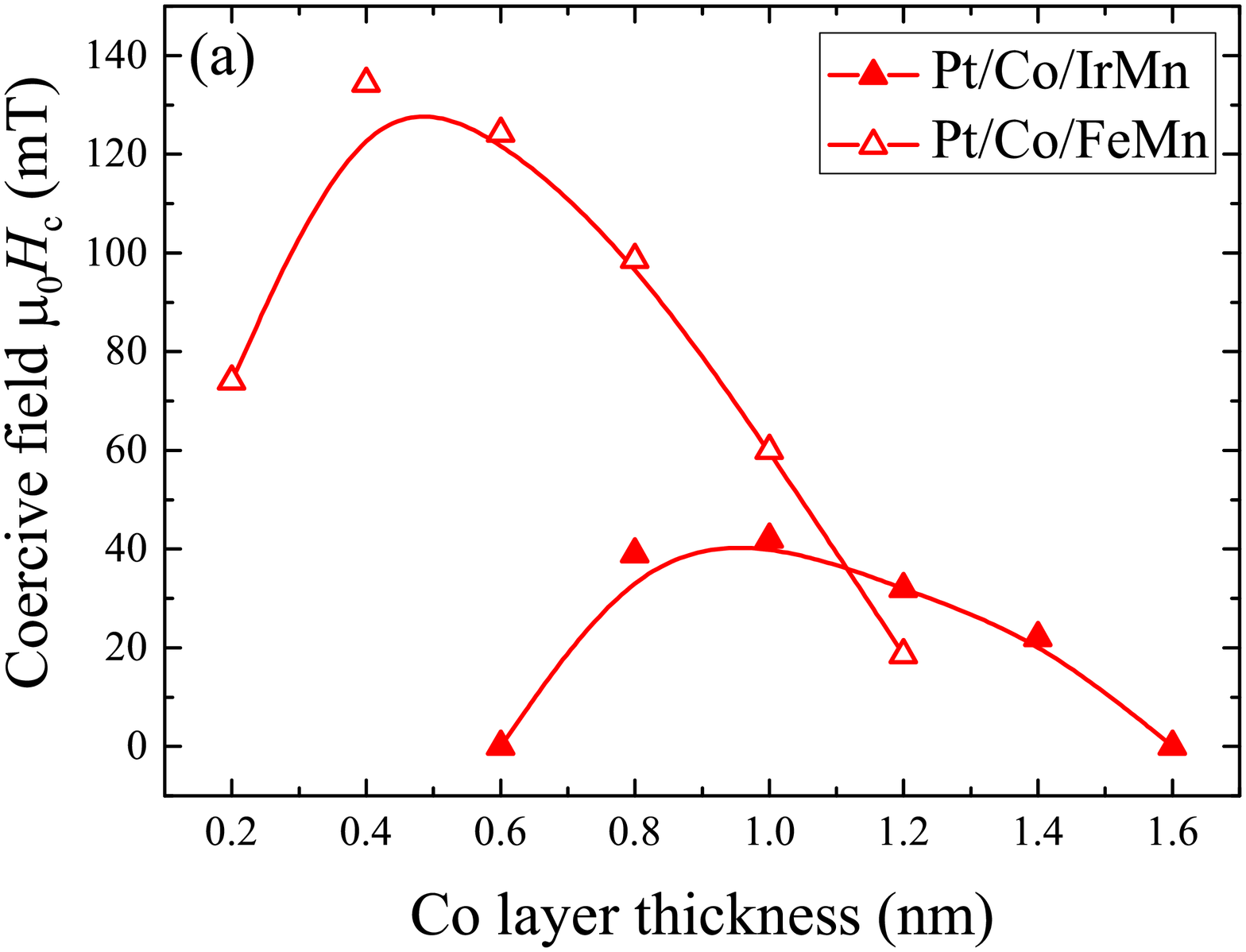}
\par\end{centering}
\begin{centering}
\includegraphics[scale=0.32]{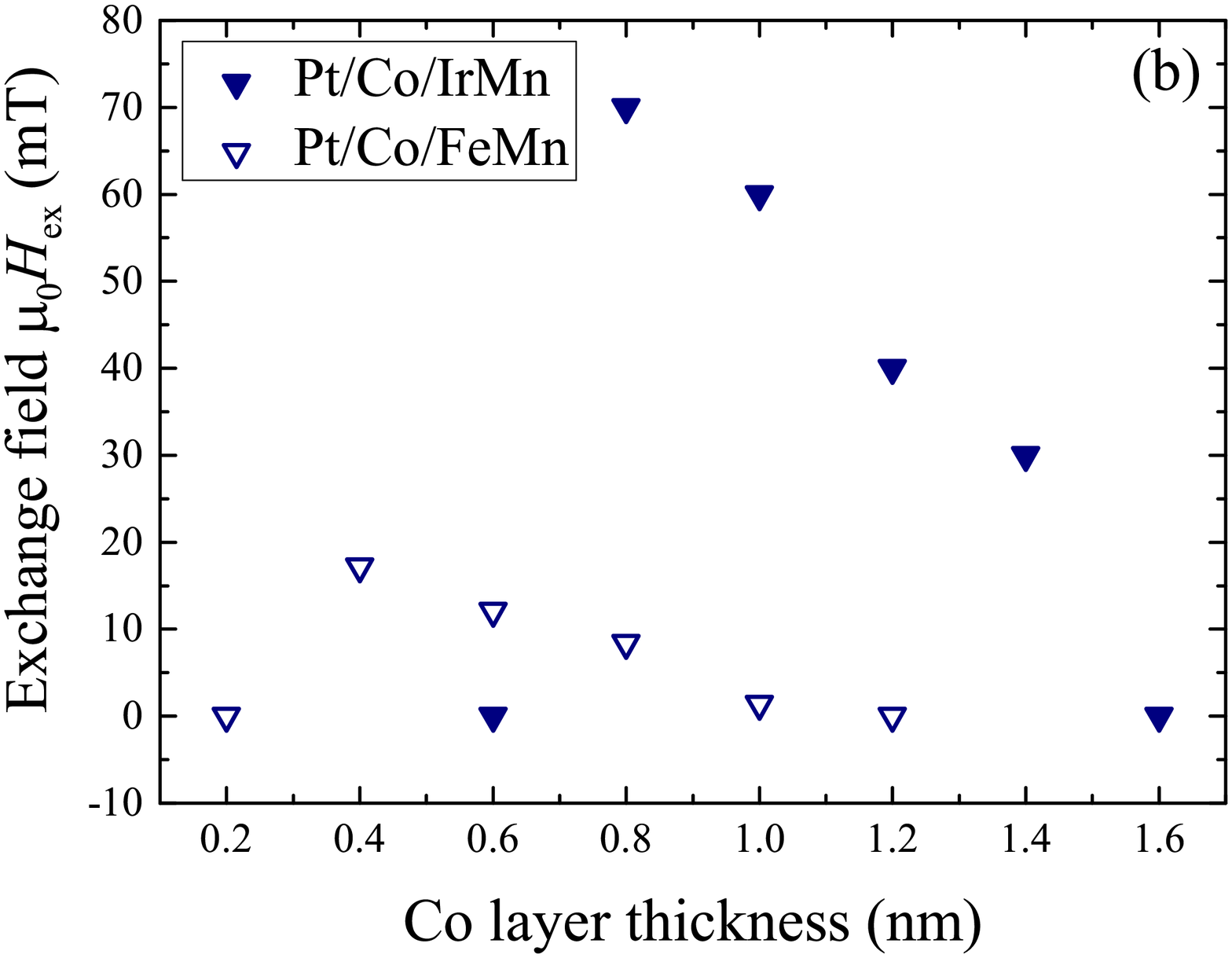}
\par\end{centering}
\caption{\label{fig:CoThicknessOpt}Coercivity and exchange bias for the systems
of Pt(2 nm)/Co(\emph{t}\protect\textsubscript{Co})/IrMn(5 nm) and
Pt(2 nm)/Co(\emph{t}\protect\textsubscript{Co})/FeMn(4 nm): (a) Coercivity
$\mathrm{\mu}_{0}H_{\mathrm{c}}$ as a function of Co layer thickness
\emph{t}\protect\textsubscript{Co} from which the optimum thickness
is chosen to be 1 nm for the IrMn, and 0.6 nm for the FeMn system.
The solid lines are guides to the eye. (b) Exchange bias field as
a function of Co layer thickness.}
\end{figure}

The systems exhibit a uniaxial magnetic anisotropy perpendicular to
the plane of the sample. To ensure this, an optimum thickness of Co
layer was chosen by systematically varying the thickness for each
system. The AFM layer thicknesses were kept constant at 5 nm for IrMn
and at 4 nm for FeMn at which the respective systems exhibit an EB
field at room temperature. Fig. \ref{fig:CoThicknessOpt}(a) shows
coercive fields, obtained from polar MOKE hysteresis loops, as a function
of Co layer thickness. PMA could be achieved for a range of Co thicknesses:
0.8-1.4 nm for the IrMn system, and 0.4-1 nm for the FeMn system.
Outside this range the sample magnetization lies in-plane. We chose
the working Co thickness to be $t_{\mathrm{Co}}=1$ nm for the IrMn
system because at this thickness the system exhibits a large coercivity
and thus provides a stable perpendicular magnetization. For the same
reason, we chose $t_{\mathrm{Co}}=0.6$ nm for the FeMn system. The
magnetization in the Co layer sets the pinning direction of the IrMn
or FeMn layer resulting in the PEB. The exchange field decreases while
the Co layer thickness increases (Fig. \ref{fig:CoThicknessOpt}(b))
for both systems, in the range where the perpendicular anisotropy
is dominant. The PEB is present in the samples in the as-grown state
and does not require any post-growth processing. 

\section{Exchange bias and domain morphology in Pt/Co/IrMn}

\subsection{Magnetic properties}

\begin{figure}
\begin{centering}
\includegraphics[scale=0.32]{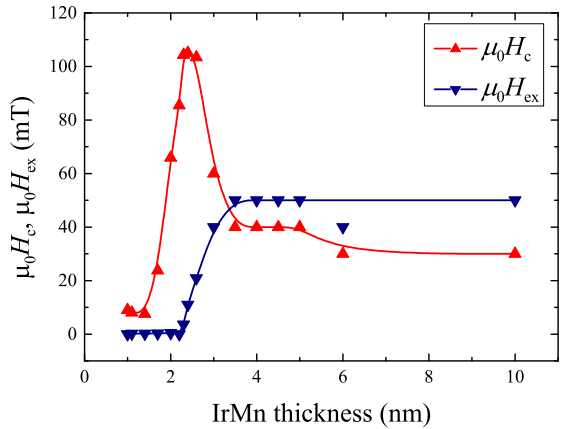}
\par\end{centering}
\caption{IrMn layer thickness dependence of the exchange bias field \emph{$\mu_{0}H_{\mathrm{ex}}$}
(blue down triangles) and coercive field \emph{$\mu_{0}H_{\mathrm{c}}$}
(red up triangles) for Pt(2 nm)/Co(1 nm)/IrMn(\emph{t}\protect\textsubscript{IrMn}).
The solid lines are guides to the eye. Onset of exchange bias occurs
at \ensuremath{\approx}2.3 nm of IrMn, at which point the coercivity
peaks.\label{fig:IrMnLayerDep} }
\end{figure}

To investigate how the exchange coupling at the FM-AFM interface modifies
the domain texture we vary the AFM layer thickness, which dictates
the spin order. We first concentrate on the IrMn system of Pt(2
nm)/Co(1 nm)/Ir\textsubscript{20}Mn\textsubscript{80}(\emph{t}\textsubscript{IrMn}),
where the IrMn layer was varied from 1 to 10 nm. A summary of coercive
fields ($H_{\mathrm{c}}$) and exchange bias fields ($H_{\mathrm{ex}}$)
is shown in Fig. \ref{fig:IrMnLayerDep}. These magnetic properties
were extracted from hysteresis loops measured by polar magneto-optic
Kerr effect (MOKE) magnetometry. $H_{\mathrm{c}}$ is half the difference
between the two switching fields, while $H_{\mathrm{ex}}$ is the
shift of the loop from $H=0$.

The onset of EB occurs at \ensuremath{\approx}2.3 nm of IrMn. $H_{\mathrm{ex}}$
rises steadily and stabilizes at $\mu_{0}H_{\mathrm{ex}}=50$ mT from
3.5 nm onward. The coercive field peaks at the same 2.3 nm of IrMn
at which the exchange field starts to develop. After peaking it gradually
drops and settles to a saturation value of $\mu_{0}H_{\mathrm{c}}=40$
mT at the same thickness of 3.5 nm at which the exchange field stabilizes.
The trends closely match with those that were reported for
similar systems with in-plane magnetization \cite{ali2003onset}.
The initial increase in coercivity occurs at the onset of the AFM
phase of IrMn and start of coupling with the Co layer. At 1 nm layer
thickness the IrMn is a paramagnet. As the thickness is increased,
the AFM phase sets in and there is an exchange interaction at the
Co/IrMn interface. The beginning of this phase transition is marked
by the increase in coercivity at \ensuremath{\approx}1.7 nm of IrMn.
As the Co layer is rotated, it also drags the spins of the IrMn layer
along with it, causing an enhancement in coercivity. The Co spins are
able to drag the IrMn spins because the volume anisotropy energy (\emph{K}\textsubscript{AFM})
of the AFM layer is smaller than the exchange energy (\emph{J}\textsubscript{FM-AFM})
at the interface between the FM and the AFM layers (\emph{K}\textsubscript{AFM}$<$\emph{J}\textsubscript{FM-AFM}).
As the IrMn thickness is increased further, \emph{K}\textsubscript{AFM}
becomes larger, resulting in further enhancement in coercivity until
a critical thickness of \ensuremath{\approx}2.3 nm is reached when
it is no longer energetically favorable for the Co layer to drag the
coupled IrMn spins. In other words, from this critical thickness onward,
\emph{K}\textsubscript{AFM} is large enough to resist the torque
from the FM Co layer (\emph{K}\textsubscript{AFM}$>$\emph{J}\textsubscript{FM-AFM}).
Thus, the coercivity gradually decreases while the exchange
field starts to increase.

\subsection{Domain morphology}

\begin{figure*}
\begin{centering}
\includegraphics[scale=0.90]{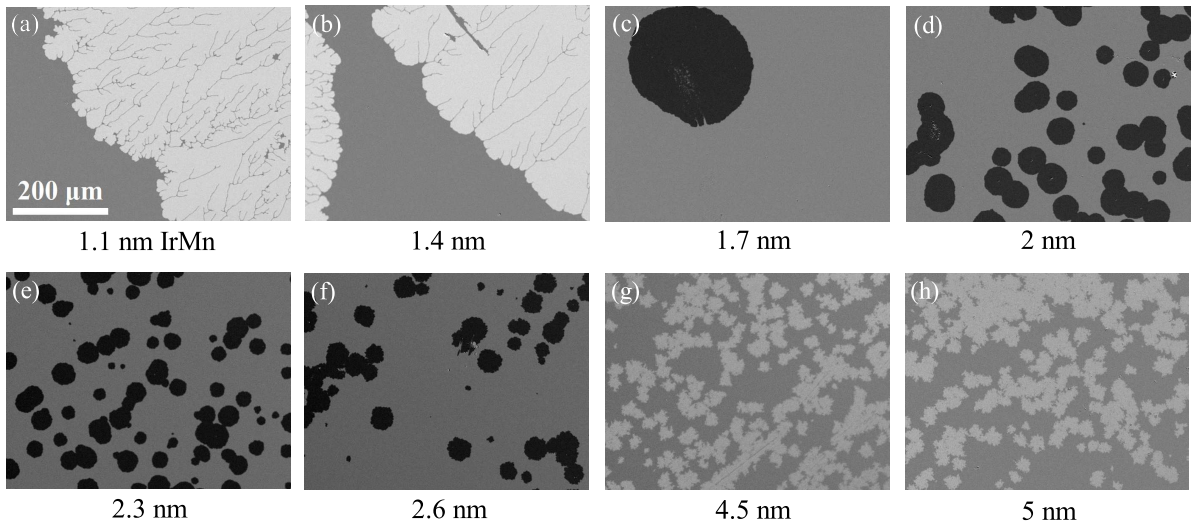}
\par\end{centering}
\caption{Kerr microscope difference images showing the propagation of magnetic
domains after $\mu_{0}H_{z}$ field pulses, which were applied for
a few seconds, and ranged from 5 to 100 mT depending on the coercivity
of the sample. The bright/dark regions (relative to grey) represent
the areas swept out by DWs during the field pulse. The domain texture
changes significantly as a function of IrMn layer thickness.\label{fig:IrMnDomainImages}
At low IrMn thicknesses (paramagnet phase), the domains are large
(a-b). At 1.7 nm thickness, the AFM phase sets in and bubble domains are
nucleated from isolated pinning sites (c). An increase in nucleation
density occurs (d) at a slightly thicker IrMn layer. Further
increasing the thickness the DWs become rough (e-f) due to the onset
of EB. Eventually, the DWs become rougher when the EB stabilizes (g-h).}
\end{figure*}

The anisotropy energy of the AFM, and consequently the FM-AFM inter-layer
coupling has a profound effect on the domain morphology. Fig. \ref{fig:IrMnDomainImages}
shows the variation in domain structure as a function of IrMn layer
thickness. The domains were imaged using a wide-field Kerr microscope
in the polar configuration, at which it is sensitive to out-of-plane
(OOP) magnetization \cite{hubert2008magnetic}. Images captured before and after the application
of an OOP field were subtracted, resulting in these difference images.
At low IrMn thicknesses, when it is in the paramagnet phase, the domains
are large and threaded with disconnected networks of unreversed narrow
domains (Fig. \ref{fig:IrMnDomainImages}(a-b)). These narrow domains
form as a DW gets pinned at a defect and bends around it. These domains
continue to exist since they are bounded by homochiral DWs, which
require large fields to annihilate because they have the same chirality
due to the DMI, and thus present a topological energy barrier \cite{benitez2015magnetic}.
At \ensuremath{\approx}1.7 nm of IrMn, coupling is initiated due to
AFM ordering and the domain morphology changes significantly. Now
bubble domains form with relatively smooth DWs instead of
the networks of narrow domains as the DWs are no longer pinned
at defect sites; Fig. \ref{fig:IrMnDomainImages}(c). This is due
to the application of relatively larger fields to nucleate domains
and propagate DWs, because of the increase in coercivity of the film
brought about by the FM-AFM coupling. It is also because of this enhancement
in coercivity that the nucleation density increases significantly
with IrMn thickness as even larger fields are now necessary to nucleate
domains. This is depicted in Fig. \ref{fig:IrMnDomainImages}(d-e).
At the critical thickness of \ensuremath{\approx}2.3 nm, the EB field
starts to set in and the DWs start to become rough (Fig. \ref{fig:IrMnDomainImages}(e-f))
due to enhanced pinning brought about by the EB, which complicates
the spin structure and increases disorder. Eventually, the DWs become
even rougher when the system exhibits a stable EB field from 3.5 nm
onward as the anisotropy energy of the AFM layer becomes robust; Fig.
\ref{fig:IrMnDomainImages}(g-h).

\subsection{Investigation of paramagnetic behavior}

\begin{figure}
\begin{centering}
\includegraphics[scale=0.5]{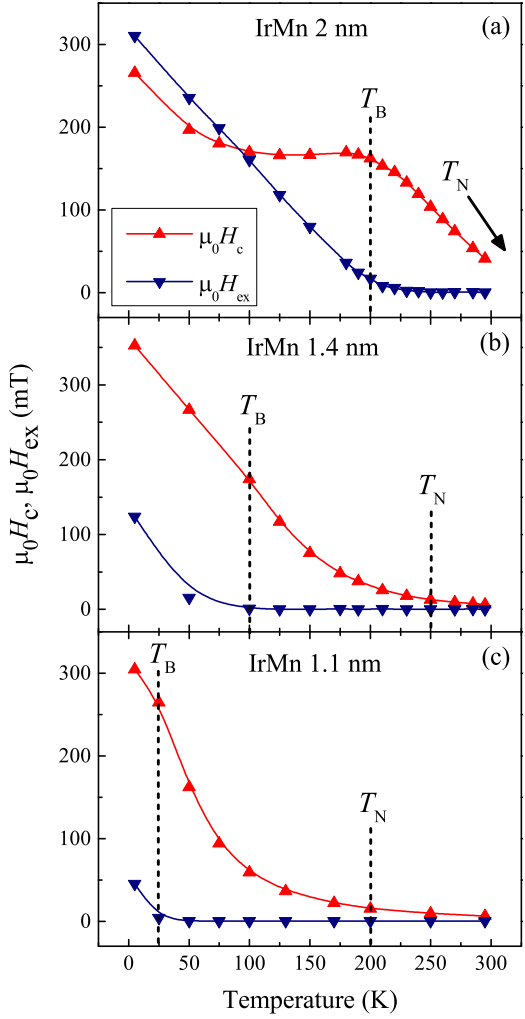}
\par\end{centering}
\caption{Temperature dependence of the coercivity \emph{$\mu_{0}H_{\mathrm{c}}$}
(red up triangles) and exchange bias field \emph{$\mu_{0}H_{\mathrm{ex}}$}
(blue down triangles) for selected IrMn thicknesses. The solid lines
are guides to the eye. Samples were initially cooled to 5 K in a 200
mT field. The blocking ($T_{\mathrm{B}}$) and N{\'e}el temperatures ($T_{\mathrm{N}}$)
of the three samples are indicated by vertical dashed lines. At 2
nm IrMn (a) $T_{\mathrm{B}}=200$ K, while $T_{\mathrm{N}}$ lies
just above room temperature. $T_{\mathrm{B}}$ and $T_{\mathrm{N}}$
move to lower temperatures as the IrMn thickness is decreased
to 1.4 nm (b) and then to 1.1 nm (c).\label{fig:TempDepOfIrMn}}
\end{figure}

We confirm paramagnetic behavior at low thicknesses by investigating
the N{\'e}el temperature ($T_{\mathrm{N}}$) and blocking temperature
($T_{\mathrm{B}}$) of the IrMn layer. We do the magnetic characterization
in a 2-300 K vibrating sample magnetometer (VSM). For this, we initially
cooled the sample from room temperature to 5 K while applying a static
perpendicular field of 200 mT, which is large enough to completely
saturate the Co layer. Then temperature dependence measurements were
done as a series of hysteresis loops at increasing temperatures. Four
repeats of field sweep were performed at each temperature to take
into account the training effect \cite{ali2003onset} and the last
loop was used for characterization. Fig. \ref{fig:TempDepOfIrMn}
shows the temperature dependence of $H_{\mathrm{c}}$ (red up triangles)
and $H_{\mathrm{ex}}$ (blue down triangles) for three of the smaller
thicknesses of IrMn (2, 1.4 and 1.1 nm). The thickness of the other
layers are as previously. $H_{\mathrm{ex}}$ of all the samples falls
with temperature and goes to zero at $T_{\mathrm{B}}$. This
is the temperature below which the AFM domains are stable and non-reversible.
$H_{\mathrm{c}}$ also shows a downward trend with temperature until
$T_{\mathrm{N}}$, at which temperature there is no AFM ordering and
the value of $H_{\mathrm{c}}$ is intrinsic to that of the Co layer.
At 2 nm IrMn $T_{\mathrm{B}}=200$ K, while $T_{\mathrm{N}}$ lies
just above room temperature (Fig. \ref{fig:IrMnLayerDep}(a)). With
decreasing IrMn thickness, both $T_{\mathrm{B}}$ and $T_{\mathrm{N}}$
shift down the temperature scale, as depicted in Fig. \ref{fig:IrMnLayerDep}(b)
and Fig. \ref{fig:IrMnLayerDep}(c) for 1.4 and 1.1 nm of IrMn, respectively.
This demonstrates that at low thicknesses, the IrMn layer is, indeed,
in the paramagnetic phase and can be made to transit to the AFM phase
just by cooling under a field. These experiments also show that $T_{\mathrm{B}}$
and $T_{\mathrm{N}}$ can be tuned easily in this system by controlling
the anisotropy energy of the IrMn layer via its thickness.

\section{Exchange bias and domain morphology in Pt/Co/FeMn}

\begin{figure}
\begin{centering}
\includegraphics[scale=0.4]{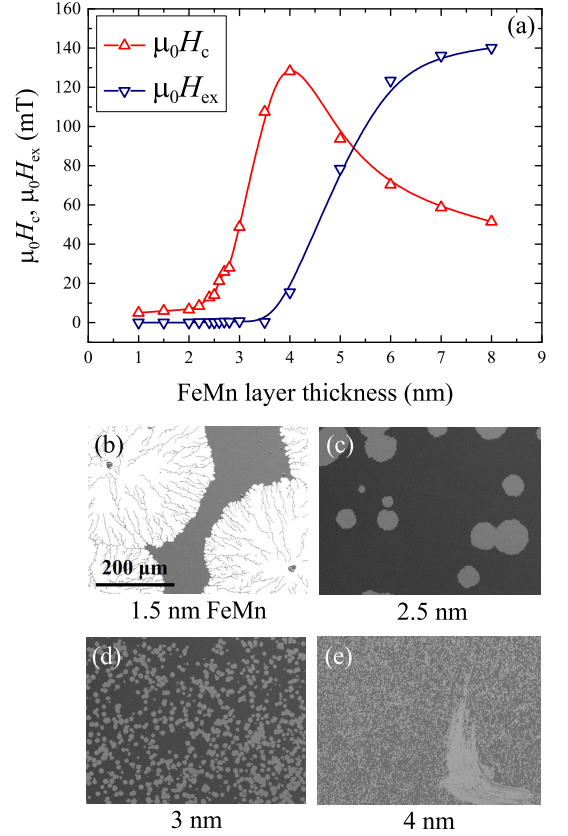}
\par\end{centering}
\caption{(a) FeMn layer thickness dependence of the exchange field \emph{H}\protect\textsubscript{ex}
(blue down triangles) and coercive field \emph{H}\protect\textsubscript{c}
(red up triangles). The solid lines are guides to the eye. (b-e) Kerr
microscope difference images showing the changes in domain structure
as a function of FeMn layer thickness.\label{fig:FeMnSummary}}
\end{figure}

The interaction mechanism, and subsequently the change in the domain
structure is similar in the system of Pt(2 nm)/Co(0.6 nm)/Fe\textsubscript{50}Mn\textsubscript{50}(\emph{t}\textsubscript{FeMn}).
Fig. \ref{fig:FeMnSummary} summarizes the dependence of $H_{\mathrm{c}}$,
$H_{\mathrm{ex}}$, and domain morphology on the FeMn layer thickness.
The onset of EB occurs at \ensuremath{\approx}4 nm of FeMn (Fig. \ref{fig:FeMnSummary}(a))
with a peak in coercivity. At \ensuremath{\approx}1.5 nm of thickness,
the FeMn layer is in the paramagnet phase and the sample exhibit
large domains containing network-like features, similar to the IrMn
system; Fig. \ref{fig:FeMnSummary}(b). At \ensuremath{\approx}2.5
nm of FeMn the AFM order sets in, causing an enhancement in coercivity
due to FM-AFM inter-layer coupling. We now see the formation of
bubble domains with smooth DWs (Fig. \ref{fig:FeMnSummary}(c)) and
without the network-like features. An increase of the FeMn layer causes
the coercivity to increase further resulting in a substantial increases
in nucleation density; Fig. \ref{fig:FeMnSummary}(d-e).

\section{Dzyaloshinskii-Moriya Interaction}

\begin{figure}
\begin{centering}
\includegraphics[scale=0.34]{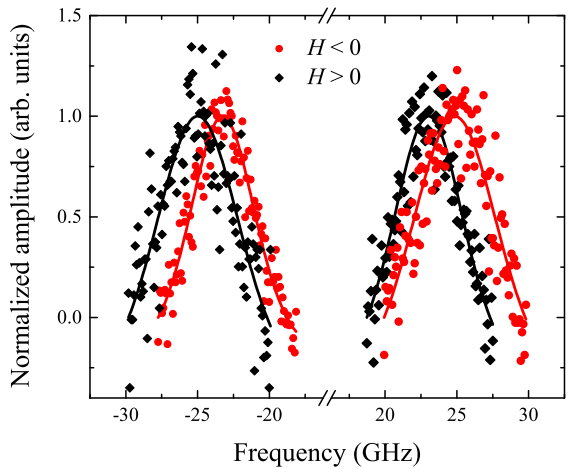}
\par\end{centering}
\caption{BLS spectra of the Damon-Eshbach spinwave modes with a wave vector of
$k=16.7\ \mathrm{\mu m^{-1}}$ for a sample of Pt(2 nm)/Co(1 nm)/IrMn(2.4
nm) with perpendicular magnetic anisotropy. The solid lines are fits
of the data using the transmission function of the tandem multi-pass
interferometer in the BLS spectrometer, $((f-f_{0})^{2}-\Delta f^{2})^{-6}$,
where $f$ is the frequency, $f_{0}$ is the resonance frequency,
and $\Delta f$ is the linewidth. The fits are normalized.\label{fig:BLSPlot}}

\end{figure}

We turn our attention to quantifying the DMI in these systems, and
infer the DW spin texture. The DMI originates at the interface where
adjacent spins of the FM undergo a chiral twist due to the exchange
interaction mediated by an atom, with a large spin-orbit coupling,
from the adjacent HM \cite{emori2013current,ryu2013chiral} or AFM
\cite{ma2017dzyaloshinskii} layer. The DMI acts locally on a DW
manifesting as an effective in-plane field. This DMI field stabilizes
the DW in a chiral N{\'e}el configuration \cite{chen2013novel,thiaville2012dynamics}
by converting it from the magnetostatically favored Bloch configuration.
We measured the DMI using Brillouin light scattering (BLS) spectroscopy 
\cite{nembach2015linear,moon2013spin,di2015direct}. In this
method, we utilize the non-reciprocity of the DMI-induced frequency-shift
and measure the Damon-Eshbach spin-wave frequencies for both field
polarities. The frequency shift is then given by

\[
\Delta f=\left|\frac{g\mu_{\mathrm{B}}}{h}\right|\frac{2D_{\mathrm{eff}}}{M_{\mathrm{s}}}k=\left|\frac{g\mu_{\mathrm{B}}}{h}\right|\frac{2D_{\mathrm{s}}}{M_{\mathrm{s}}t_{\mathrm{FM}}}k,
\]
where \emph{D}\textsubscript{eff}\emph{ }is the volumetric DMI constant
that determines the sign and magnitude of the DMI vector, $g$ is
the spectroscopic $g$-factor taken to be 2.14 \cite{schoen2017magnetic},
\emph{M}\textsubscript{s} is the saturation magnetization, \textbf{k}
(with magnitude \emph{k}) is the wavevector of the spin waves, $\mu_{\mathrm{B}}$
is the Bohr magneton, and \emph{h} is Planck constant. The sign
of the frequency-shift depends on the direction of the magnetization
and the propagation direction of the spin-waves. In the last equality,
\emph{D}\textsubscript{s}\emph{ }is the interfacial DMI parameter,
which represents the DMI contribution from the top and bottom interfaces
($D_{\mathrm{s}}=D_{\mathrm{eff}}t_{\mathrm{FM}},$ where $t_{\mathrm{FM}}$
is the FM layer thickness). Thus, \emph{D}\textsubscript{s} should
be independent of the FM layer thickness, if we consider the DMI to
be a truly interfacial effect. Figure \ref{fig:BLSPlot} shows representative
BLS spectra for a sample of Pt(2 nm)/Co(1 nm)/IrMn(2.4 nm), where
shifts in Stokes (negative frequencies) and anti-Stokes (positive
frequencies) peaks are evident, corresponding to $\Delta f=-0.90\pm0.05$
GHz.

The saturation magnetization is measured by a superconducting quantum
interference device (SQUID) VSM. For Pt/Co/IrMn, $M_{\mathrm{s}}=(1.36\pm0.05)\times10^{6}$
A/m, which is similar to the value for bulk Co. For Pt/Co/FeMn, $M_{\mathrm{s}}=(2.33\pm0.05)\times10^{6}$
A/m. The high value in this case may be due to the formation of
a monolayer of Fe at the Co/FeMn interface \cite{antel1999spin}, 
which contributes to the total moment. Thus, to account for this,
we increase the effective volume of the FM layer by including a monolayer
of Fe and we arrive at a value of $M_{\mathrm{s}}=(1.57\pm0.04)\times10^{6}$
A/m.

\begin{table*}[t]
\caption{\label{tab:Comparison}Comparison of the DMI at different phases of
the AFM layers of the two investigated exchange coupled systems. The
numbers in parentheses represent the nominal layer thickness in nanometers.
The dominant contribution to the uncertainty in \emph{D}\protect\textsubscript{eff}
is the uncertainty in \emph{M}\protect\textsubscript{s}, whereas for
\emph{D}\protect\textsubscript{s}, there is also a contribution from
the uncertainty in \emph{t}\protect\textsubscript{FM}, leading to
a larger experimental error. For the FeMn system, the contribution
of a monolayer of Fe to the volume of the FM layer and the \emph{M}\protect\textsubscript{s} 
was taken into account when calculating \emph{D}\protect\textsubscript{s} (as outlined in the text).}
\centering{}%
\begin{tabular}{ccccc}
\hline
\hline 
Sample & AFM layer spin order & Exchange bias & \emph{D}\textsubscript{eff}\emph{ }(mJ/m\textsuperscript{2}) & \emph{D}\textsubscript{s} (pJ/m)\tabularnewline
\hline 
(a) Pt(2)/Co(1)/IrMn(1.1) & Paramagnetic & No & $-1.14\pm0.05$ & $-1.14\pm0.13$\tabularnewline
(b) Pt(2)/Co(1)/IrMn(1.7) & Antiferromagnetic & No & $-1.14\pm0.05$ & $-1.14\pm0.12$\tabularnewline
(c) Pt(2)/Co(1)/IrMn(2.4) & Antiferromagnetic & Yes & $-1.22\pm0.08$ & $-1.22\pm0.15$\tabularnewline
(d) Pt(2)/Co(1)/IrMn(5) & Antiferromagnetic & Yes & $-1.11\pm0.12$ & $-1.11\pm0.16$\tabularnewline
\hline 
(e) Pt(2)/Co(0.6)/FeMn(1) & Paramagnetic & No & $-1.50\pm0.08$ & $-1.35\pm0.17$\tabularnewline
(f) Pt(2)/Co(0.6)/FeMn(2.6) & Antiferromagnetic & No & $-1.44\pm0.08$ & $-1.30\pm0.16$\tabularnewline 
\hline
\hline
\end{tabular}
\end{table*}

The magnitude of the DMI of the two systems at different AFM layer
thicknesses are summarized in Table \ref{tab:Comparison}. The experiments
show that the spin order of the AFM layers does not affect the DMI.
To assess this, we measured the DMI of the IrMn system at four critical
thicknesses of the IrMn layer, which correspond to the paramagnet
phase (sample (a)) with no spin order, the AFM phase at the point
of paramagnet-to-AFM phase transition (sample (b)) with no EB, the
AFM phase at peak coercivity (sample (c)) when the EB starts to set
in, and the AFM phase with a large EB (sample (d)) brought about by
a larger anisotropy energy of the AFM layer as being thicker. The
magnitude of the DMI remains the same in all four cases, from which
we conclude that neither the spin order of the AFM layer nor the
EB play a role in the mechanism of the DMI in this system. The same 
behavior occurs in the FeMn system. The DMI of the system when
the FeMn layer is paramagnetic (sample (e)) is the same when it is
antiferromagnetic (sample (f)). It was not possible to measure the DMI 
for samples with a large EB: a thick FeMn layer reduces the backscattered 
signal, and the very thin Co layer has large linewidth because of spin-pumping 
and two-magnon scattering due to the presence of the AFM layer. Both 
systems possess left-handed chirality (counter-clockwise).

According to the three-site model of Fert and Levy \cite{fert1980role},
a DM-type interaction occurs when an impurity atom, due to the spin-orbit
coupling (SOC) of its conduction electrons, mediates an exchange interaction
between two magnetic atoms. The SOC constant does not depend on the
spin state but rather on the atomic number. Hence, the DMI is not
influenced by the spin order of the AFM layer. Our experiment is in
accordance with this model.

Our measurements show that the DMI remains unchanged with and without
the presence of an EB field. However, we do note that the change in
DMI at the CoFeB/IrMn interface as the IrMn film thickness is increased
from 1 to 8 nm is of the order of 0.1 mJ/m\textsuperscript{2} \cite{ma2017dzyaloshinskii}.
We cannot rule out a similar change in our Co/IrMn system because such
a value falls within our experimental uncertainty. The relatively
large error for sample (d) is again due to the presence of a thick
AFM IrMn layer.

The FeMn system has a different \emph{D}\textsubscript{s} than the
IrMn system. This indicates that the DMI is different at the two interfaces
of Co/IrMn and Co/FeMn, as the Pt/Co interface is common to both.
This could be expected due to the difference in Mn concentration for
the two AFMs (Mn atoms mostly contribute to the DMI \cite{ma2017dzyaloshinskii}).
This is the case for a ``clean'' Co/FeMn interface. However, the
formation of an Fe layer at the Co/FeMn interface could mean that
we need to consider the contribution of both ferromagnetic Fe and
Co to the DMI. Furthermore, due to intermixing, CoMn, which is antiferromagnetic,
could also play a role in the generation of the DMI. We also point
out here that although the quantity \emph{D}\textsubscript{s} is
normalized with respect to the FM layer thickness, Nembach et al.
\cite{nembach2015linear} has shown a non-trivial relationship between
the two and suggested that FM thickness could also change the interfacial
DMI.

\section{Conclusion}

We investigated interfacial mechanisms in exchange-coupled systems
of Pt/Co/IrMn and Pt/Co/FeMn exhibiting perpendicular magnetic anisotropy
and perpendicular exchange bias. We control the spin order of the
antiferromagnet layers by varying the thicknesses. We study the
changes in the magnetic domain morphology by magneto-optical imaging,
and the interfacial Dzyaloshinskii-Moriya interaction by Brillouin
light scattering spectroscopy. We demonstrate that the domain structure
in these systems is dictated by the AFM N{\'e}el order. The domain texture
changes from large domains with unreversed networks to isolated bubbles
with smooth DWs at the onset of AFM order. The DWs become rough
due to pinning as the exchange bias field develops. These changes
are linked to the anisotropy energy of the AFM layer and the FM-AFM
inter-layer exchange coupling. The DMI is not influenced
by the AFM spin order within experimental uncertainty, in agreement with theory.

\begin{acknowledgments}
This work was funded by the European Community under the Marie-Curie
Seventh Framework program - ITN \textquotedblleft WALL\textquotedblright{}
(Grant No. 608031). Equipment funding was provided by U.K. EPSRC;
Grant No. EP/K003127/1 for the Kerr microscope, and Grant No. EP/K00512X/1
for the SQUID VSM. Support from European Union grant MAGicSky No.
FET-Open-665095.103 is gratefully acknowledged. The authors thank G. Durin 
for helpful discussions, and A. Westerman, K. Zeissler and T. Moorsom 
for assisting with experiments.
\end{acknowledgments}

\bibliography{References_Khan}

\end{document}